\documentclass[nofootinbib,aps,amsmath,amssymb,a4paper,prd,twocolumn]{revtex4-1}
\usepackage[american]{babel}
\usepackage[utf8]{inputenc}
\usepackage[T1]{fontenc}

\usepackage{graphicx}
\usepackage{color}

\usepackage{units}
\usepackage{hyperref}
\hypersetup{
   pdfnewwindow=true,     % links in new window
   colorlinks=true,       % false: boxed links; true: colored links
   linkcolor=black,       % color of internal links
   citecolor=blue,        % color of links to bibliography
   filecolor=black,       % color of file links
   urlcolor=blue          % color of external links
}

\begin{document}

%%%%%%%%%%%%%%%%%%%%%%%%%%%%%%%%%%%%%%%%%%%%%%%%%%%%%%%%%%%%%%%%%%%%%%%%%%%%%%%
% Title start
%%%%%%%%%%%%%%%%%%%%%%%%%%%%%%%%%%%%%%%%%%%%%%%%%%%%%%%%%%%%%%%%%%%%%%%%%%%%%%%
\title{\Large \textbf{Gamma Lines from Majorana Dark Matter}}
\author{Michael \surname{Duerr}}
%\email{michael.duerr@mpi-hd.mpg.de}
\author{Pavel \surname{Fileviez P\'erez}}
%\email{fileviez@mpi-hd.mpg.de}
\author{Juri \surname{Smirnov}}
%\email{juri.smirnov@mpi-hd.mpg.de}
\affiliation{\small{
Particle and Astroparticle Physics Division,\\
Max-Planck-Institut f\"ur Kernphysik,\\
Saupfercheckweg 1, 69117 Heidelberg, Germany}
}
%%%%%%%%%%%%%%%%%%%%%%%%%%%%%%%%%%%%%%%%%%%%%%%%%%%%%%%%%%%%%%%%%%%%%%%%%%%%%%%
%%%%%%%%%%%%%%%%%%%%%%%%%%%%%%%%%%%%%%%%%%%%%%%%%%%%%%%%%%%%%%%%%%%%%%%%%%%%%%%
\begin{abstract}
We discuss simple models which predict the existence of significant gamma-ray fluxes from dark matter annihilation. 
In this context the dark matter candidate is a Majorana fermion with velocity-suppressed tree-level annihilation into Standard Model fermions but unsuppressed annihilation into photons. These gamma lines can easily be distinguished from the continuum and provide a possibility to test these models.  
\end{abstract}
%%%%%%%%%%%%%%%%%%%%%%%%%%%%%%%%%%%%%%%%%%%%%%%%%%%%%%%%%%%%%%%%%%%%%%%%%%%%%%%
%%%%%%%%%%%%%%%%%%%%%%%%%%%%%%%%%%%%%%%%%%%%%%%%%%%%%%%%%%%%%%%%%%%%%%%%%%%%%%%

\maketitle

%%%%%%%%%%%%%%%%%%%%%%%%%%%%%%%%%%%%%%%%%%%%%%%%%%%%%%%%%%%%%%%%%%%%%%%%%%%%%%%
\section{Introduction}
%%%%%%%%%%%%%%%%%%%%%%%%%%%%%%%%%%%%%%%%%%%%%%%%%%%%%%%%%%%%%%%%%%%%%%%%%%%%%%%
The nature of the dark matter~(DM) component of our Universe is a long standing mystery and one of the most challenging questions in fundamental physics. It has been argued for a long time that the observation of monochromatic gamma-ray lines from dark matter annihilation would be a ``smoking gun'' signature for the particle nature of dark matter. This statement is correct in the sense that it is unlikely that this signature could have an astrophysical origin. However, the question whether in any DM model such a feature is generic is much more subtle. 
 
\begin{figure}[b]
\centering
\includegraphics[width=0.7\linewidth]{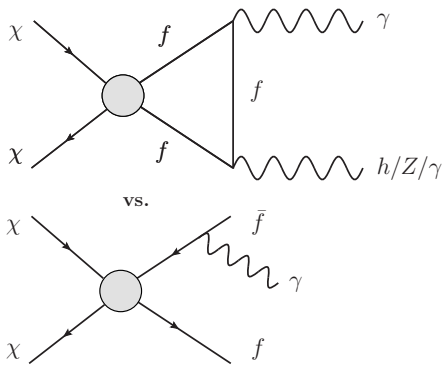}
\caption{DM annihilation with photons in the final state. 
\label{fig:effectiveCoupling}}
\end{figure}

In order to clearly identify a gamma line from dark matter annihilation the contribution of the dark matter annihilation to final state radiation~(FSR) and any annihilation channel contributing to the gamma continuum have to be small and the annihilation into one or more of the final states $\gamma \gamma$, $h \gamma$ and $Z \gamma$ has to be large. Typically, it is difficult to have such a scenario because the final state radiation occurs at tree level while the gamma lines are possible only at the quantum level, see Fig.~\ref{fig:effectiveCoupling}. For a review on DM annihilation into gamma rays see Ref.~\cite{Bringmann:2012ez}.

In the simplest dark matter models where the dark matter relic density is defined by the annihilation into the Standard Model (SM) fermions one generically expects to have a large contribution to final state radiation. 
For example, if the dark matter candidate has spin zero and annihilates through the Higgs into SM fermions one can explain the relic density through thermal freeze-out but 
at the same time the contribution to final state radiation tends to be large. In this case one can find regions in the parameter space where the FSR is suppressed\footnote{For examples, see Refs.~\cite{Gustafsson:2007pc,Duerr:2015mva,Duerr:2015aka}.} but generically it is difficult to see a gamma line.

The case of Dirac dark matter is similar to the scalar dark matter scenario. For example, if the Dirac DM candidate is charged under an Abelian gauge force one can explain the relic density through the interactions with the new gauge boson. Typically, the tree level annihilation is not suppressed and therefore one expects to have a large contribution to final state radiation.  

In the context of the Minimal Supersymmetric Standard Model the lightest neutralino, which is a Majorana fermion, can be a cold dark matter candidate~\cite{Jungman:1995df}. 
A case with interesting $\gamma$-ray phenomenology is when the neutralino is bino-like and the annihilation through the t-channel is velocity suppressed.
However, in this set-up virtual internal bremsstrahlung~(VIB) from the mediating squarks and sleptons leads to a gamma ray continuum. This contribution spoils the gamma line visibility. It has been argued in the literature that the VIB process itself can mimic a line signal, but it is only possible in a strongly tuned mass degenerate scenario. See Refs.~\cite{Giacchino:2013bta,Bringmann:2012vr,Kopp:2014tsa,Gustafsson:2007pc,Arina:2009uq,Dudas:2012pb,Jackson:2009kg,Duerr:2015wfa,Garcia-Cely:2015dda,Coogan:2015xla} for recent studies of gamma spectra in different dark matter models.

A case where one could easily have a visible gamma line corresponds to the scenario where the tree level annihilation 
of a Majorana fermion into SM fermions is velocity suppressed. Since the dark matter velocity today is very small one could hope that the contribution to final state radiation is small but the rate for gamma lines is large. In this way the velocity suppression compensates for the loop factor present in the gamma lines.

In this letter we investigate a simple scenario where the dark matter candidate is a Majorana fermion which naturally gives rise to a significant gamma line and negligible contribution to the continuum. In this way one could hope that experiments such as H.E.S.S.\ looking for gamma lines from dark matter annihilation could cut into the parameter space of this model.

%%%%%%%%%%%%%%%%%%%%%%%%%%%%%%%%%%%%%%%%%%%%%%%%%%%%%%%%%%%%%%%%%%%%%%%%%%%%%%%
\section{Majorana Dark Matter}
%%%%%%%%%%%%%%%%%%%%%%%%%%%%%%%%%%%%%%%%%%%%%%%%%%%%%%%%%%%%%%%%%%%%%%%%%%%%%%%
Let us consider a model where the dark matter is a Majorana fermion $\chi$ charged under a new local gauge symmetry. Once the new gauge boson is integrated out one could have the effective interaction
\begin{equation}\label{eq:effectiveInteractionAA}
\mathcal{L}_\text{DM} \supset  \frac{c_\text{AA}}{\Lambda^2} \bar{\chi} \gamma^\mu \gamma^5 \chi \bar{f} \gamma_\mu \gamma^5  f.
\end{equation}
Here $f$ could be a SM fermion or a new fermion needed for anomaly cancellation in models with extra gauge symmetries. 
Once we dress this operator to study the annihilation $\chi \chi \to \gamma \gamma$, one finds that this channel is not velocity suppressed.
The annihilation $\chi \chi \to h \gamma$ is suppressed when $f$ is a new fermion, because the Yukawa couplings for the new fermions inside the loop have to be small. 
In Ref.~\cite{Ishiwata:2011hr} it was shown that new chiral fermions with large Yukawa couplings to the SM Higgs significantly change the $h \to \gamma\gamma$ cross section and are therefore ruled out. Thus, new fermions needed for anomaly cancellation need to be vector-like under the SM gauge group and obtain the dominant part of their mass from a new Higgs field, not from the SM Higgs.  
The case of $\chi \chi \to Z \gamma$ is similar to the annihilation into gamma gamma. 

The Majorana fermion could have only the following interaction
\begin{equation}\label{eq:effectiveInteractionAV}
\mathcal{L}_\text{DM} \supset  \frac{c_\text{AV}}{\Lambda^2} \bar{\chi} \gamma^\mu \gamma^5 \chi \bar{f} \gamma_\mu   f.
\end{equation}
In this case the annihilation into $\gamma \gamma$ is not present and $h \gamma$ is velocity suppressed, while the annihilation into $Z \gamma$ is not velocity suppressed.
See Table~\ref{tab:operators} for the summary of the features of all possible channels leading to gamma lines. In general one could have a model where both interactions in Eqs.~\eqref{eq:effectiveInteractionAA} and \eqref{eq:effectiveInteractionAV} are present and one could hope to observe the $\gamma \gamma$ and $Z \gamma$ channels.
In what follows, we discuss a model where the dark matter annihilation into gamma lines is naturally large and could provide a possibility to test this model.

\begin{table}[b]
\caption{Effective operators for Majorana DM and expected strength of the possible gamma lines. \label{tab:operators}}
\begin{center}
\begin{tabular}{cccc}
\hline
Operator & $\gamma \gamma$ & $Z \gamma$ & $h \gamma$ \\
\hline \hline
$ \bar{\chi} \gamma^\mu \gamma^5 \chi \bar{f} \gamma_\mu \gamma^5  f$ & OK & OK & suppressed \\
%\hline
$\bar{\chi} \gamma^\mu \gamma^5 \chi \bar{f} \gamma_\mu   f$ & -- & OK & suppressed \\
\hline
\end{tabular}
\end{center}
\end{table}

%%%%%%%%%%%%%%%%%%%%%%%%%%%%%%%%%%%%%%%%%%%%%%%%%%%%%%%%%%%%%%%%%%%%%%%%%%%%%%%
\section{Baryonic Dark Matter}
%%%%%%%%%%%%%%%%%%%%%%%%%%%%%%%%%%%%%%%%%%%%%%%%%%%%%%%%%%%%%%%%%%%%%%%%%%%%%%%
We investigate the visibility of the gamma lines from dark matter annihilation in models where the baryon number is a local gauge symmetry spontaneously broken at the low scale.
These models were proposed and studied in Refs.~\cite{Duerr:2013dza,Perez:2014qfa,Duerr:2013lka,Duerr:2014wra,Ohmer:2015lxa} and are based on the gauge group
\begin{equation}
G_B = SU(3)_C \otimes SU(2)_L \otimes U(1)_Y \otimes U(1)_B.
\end{equation}
In the two realistic models~\cite{Duerr:2013dza,Perez:2014qfa} one has new fermions with baryon number which are necessary to define an anomaly-free theory.
In this context the lightest field with baryon number in the new sector is neutral and can describe the cold dark matter in the Universe. 
In both cases the dark matter candidate $\chi$ can be a Majorana fermion if the baryon number for the new fields is $\pm 3/2$. 

The interaction of the dark matter candidate with the leptophobic gauge boson $Z_B$ is given by
\begin{equation}
- \mathcal{L} \supset \frac{3}{2} g_B \bar{\chi} \gamma_\mu \gamma^5 \chi Z_B^\mu,
\end{equation} 
where $g_B$ is the new gauge coupling. 
Here we neglect for simplicity the kinetic mixing between SM hypercharge and baryon number.
Notice that all the SM quarks will feel this new force associated with the baryon number,
\begin{equation}
- \mathcal{L} \supset \frac{1}{3} g_B \bar{q} \gamma^\mu q Z^\mu_B.
\end{equation}
The new fermions $f^\prime$ needed for anomaly cancellation acquire mass from a new Higgs 
and will have axial couplings with the $Z_B$ gauge boson
\begin{equation}
- \mathcal{L} \supset  \pm \frac{3}{2} g_B \bar{f^\prime} \gamma_\mu \gamma^5  f^\prime Z^\mu_B.
\end{equation}
Notice that once the $Z_B$ is integrated out one finds the interaction in Eq.~\eqref{eq:effectiveInteractionAA} with $f$ being the new fermions needed for anomaly cancellation. 

In order to investigate the predictions for the gamma lines we will focus on the model proposed in 
Ref.~\cite{Duerr:2013dza}. In this context one has the additional fields
\begin{eqnarray}
\Psi_L & \sim & (1,2,1/2,-3/2), \ \Psi_R \sim (1,2,1/2,3/2), \nonumber \\
\eta_R & \sim & (1,1,-1,-3/2), \ {\text{and}} \ \eta_L \sim (1,1,-1,3/2), \nonumber
\end{eqnarray} 
with the transformation properties under $G_B$ given. 
These new fields together with the dark matter candidate acquire mass once 
the scalar field  $S_B \sim (1,1,0,-3)$ obtains a vacuum expectation value $\langle S_B \rangle = v_B/\sqrt{2}$ and breaks the local baryon symmetry. The $v_B$ also determines the mass of the $Z_B$ as
\begin{equation}
 M_{Z_B} =  3 g_B v_B.
\end{equation}

The relevant interactions of the new fermion fields are given by  
\begin{align}\label{eq:LagrangianMass}
- \mathcal{L} & \supset  \lambda_{\Psi} \overline{\Psi}_L {\Psi}_R S_{B} \ +   \lambda_{\eta} \overline{\eta}_R {\eta}_L S_{B} 
 +  \lambda_{\chi} \overline{\chi}_R {\chi}_L S_{B}  \ \nonumber \\ & + \frac{\lambda_1}{2} \chi_L \chi_L S_{B} + \frac{\lambda_2}{2} \chi_R \chi_R S_{B}^\dagger +  \text{h.c.}
\end{align}
Once the $U(1)_B$ symmetry is broken one has two neutral Majorana fields in the $\chi_L$--$\chi_R$ sector. The mass matrix in the $( \chi_L, \chi_L^c)$ basis is
\begin{equation}
    \label{mixingmatrix}
   \mathcal{M}= \frac{v_B}{\sqrt{2}}\left(
    \begin{array}{cc}
    \lambda_1 &  \lambda_\chi \\
     \lambda_\chi &  \lambda_2 
    \end{array}
    \right)\,.
  \end{equation}
Given the parameter constellation with the following hierarchy $\lambda_2 \gg \lambda_\chi \gg \lambda_1$ we have a seesaw in the dark matter sector leading to
$m_1 \approx  \frac{\lambda_\chi^2}{\lambda_2} \frac{v_B}{\sqrt{2}}$  and  $m_2 \approx  \lambda_2 \frac{v_B}{\sqrt{2}}$.
 Thus $\chi_1$ is the lightest particle in this sector, is stable and can be the cold dark matter candidate.
Note that the stability of the dark matter is due to a remnant $\mathcal{Z}_2$ symmetry in the new sector. Furthermore, in this set-up $\chi_2$ is significantly heavier than $\chi_1$ and co-annihilations do not play any significant role in the freeze-out process. Additionally no co-annihilations with the other neutral fermionic states occur, as we work in the regime where the Yukawa couplings of the new fermions are small and thus there is practically no mixing with the other neutral states.

In our study we focus on the scenario where the dark matter is the lightest new field in the theory. We show the 
predictions assuming that the relic density can be thermally produced through the freeze-out of our dark matter candidate. 
The relevant annihilation channel then is the annihilation to SM quarks, $\chi \chi \to Z_B^* \to \bar{q} q$, 
and the corresponding cross section is given by
\begin{equation}
 \sigma \left(\chi \chi \to \bar{q} q \right) = \frac{g_B^4 }{16 \pi s} \ \frac{\sqrt{s - 4 M_q^2} \sqrt{s - 4 M_\chi^2} \left( s + 2M_q^2 \right)}{\left[ \left(s- M_{Z_B}^2\right)^2 + \Gamma_{Z_B}^2 M_{Z_B}^2 \right]} ,
\end{equation}
where $s$ is the square of the center-of-mass energy. The s-wave contribution to this annihilation process is vanishing identically, such that it is p-wave dominated and hence velocity suppressed. The other possible annihilation channels through the two Higgs bosons have a double suppression: they are suppressed by the mixing angle between the two Higgs fields and also velocity suppressed. 

\begin{figure}[t]
\centering
\includegraphics[width=0.8\linewidth]{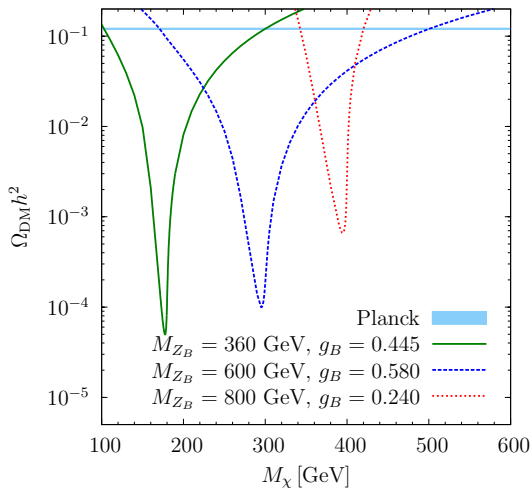}
\caption{Relic density vs.\ the dark matter mass for three choices of parameters. Note that the points in parameter space where the correct relic density is achieved correspond to the benchmark scenarios I--III. The solid light-blue line represents the currently allowed range for the DM relic density, $\Omega_\text{DM} h^2 = 0.1199 \pm 0.0027$~\cite{Ade:2013zuv}.
\label{fig:relic}}
\end{figure}

Neglecting the annihilation through the Higgs fields, in Fig.~\ref{fig:relic} we show the relic density vs.\ the dark matter mass for the benchmark scenarios in Table~\ref{tab:benchmark1}. 
The calculation of the relic density is performed in the usual way~\cite{Gondolo:1990dk}. The correct relic density can be achieved far from the resonance even if the annihilation cross 
section is velocity suppressed because the DM velocity is about $0.3 c$ at freeze out and the suppression is therefore small.

\begin{table}[b]
\caption{Benchmark scenarios. \label{tab:benchmark1}}
\begin{center}
\begin{tabular}{cccccc}
\hline
Scenario & $M_\chi$~[GeV] & $M_{Z_B}$~[GeV] & $g_B$ & $M_\eta$~[GeV] & $M_\Psi$~[GeV] \\
\hline \hline
I   & 300 & 360 & 0.445 & 300 & 600 \\
II  & 500 & 600 & 0.580 & 1000 & 1500 \\
III & 420 & 800 & 0.240 & 420 & 500 \\
\hline
\end{tabular}
\end{center}
\end{table}

In this model the predictions for direct detection depend on the mixing angle $\theta$ between the SM Higgs and the new Higgs breaking the local baryon number, since the direct detection cross section through the $Z_B$ is velocity suppressed; see  Ref.~\cite{Ohmer:2015lxa} for more details. As it has been shown there, one can have large rates when the mixing angle is close to the experimental bound, $\theta \lesssim 0.35$. 
We have checked the possible bounds from the Large Hadron Collider~(LHC) searches for this model and would like to mention that the results shown in this paper are in agreement with the current bounds. See for example Ref.~\cite{Buchmueller:2014yoa} for a detailed discussion of the LHC bounds.

%%%%%%%%%%%%%%%%%%%%%%%%%%%%%%%%%%%%%%%%%%%%%%%%%%%%%%%%%%%%%%%%%%%%%%%%%%%%%%%
\section{Visible Gamma Lines}
%%%%%%%%%%%%%%%%%%%%%%%%%%%%%%%%%%%%%%%%%%%%%%%%%%%%%%%%%%%%%%%%%%%%%%%%%%%%%%%
Recently, we have discussed the dark matter annihilation into gamma lines in simplified dark matter models~\cite{Duerr:2015wfa}. 
The calculations performed in Ref.~\cite{Duerr:2015wfa} are general and can be used for the baryonic dark matter model discussed 
here. In this model, there are three possible gamma lines,
\begin{equation}
\bar{\chi} \chi \ \to \ Z_B^* \ \to \  \gamma \gamma, \ h \gamma, \ Z \gamma,
\end{equation}
and the effective couplings $Z_B \gamma \gamma$, $Z_B h \gamma$, and $Z_B Z \gamma$ are generated by loops of 
SM quarks and/or the heavy fermions $\Psi$ and $\eta$. For relic density and for DM direct detection, the model has three free parameters: $M_\chi$, $M_{Z_B}$, and $g_B$. For the indirect detection, one has two extra parameters in the model, $M_\Psi$ and $M_\eta$, the masses of the heavy new fermions, which gives a total of five free parameters. We define three benchmark scenarios for this set of parameters in Table~\ref{tab:benchmark1}, which we will use throughout this letter. 
It is important to remind the reader that in the case of the annihilation into two gammas the energy of the gamma line is equal to the dark matter mass. 
In the case of the annihilation into $h \gamma$ and $Z \gamma$ the energy of the gamma line is
\begin{equation}
E_\gamma = M_\chi \left( 1- \frac{M_X^2}{4 M_\chi^2} \right), 
\end{equation} 
with $X=h$ or $X=Z$ for the corresponding channel.

The cross section for the annihilation to two gammas is given by
\begin{multline}\label{eq:gammagammaCrossSection}
 \sigma(\bar{\chi} \chi \to \gamma \gamma) =  \frac{81 \alpha^2}{32 \pi^3} \frac{g_B^4 M_\chi^2}{ M_{Z_B}^2 \left( M_{Z_B}^2 + \Gamma_{Z_B}^2 \right) } \\
 \times \frac{\left[  M_\Psi^2 C_0 (s;M_\Psi) - M_\eta^2 C_0 (s;M_\eta) \right]^2}{\sqrt{1 - 4 M_\chi^2/s}} ,
 \end{multline}
where $\alpha=e^2/ 4 \pi$. The scalar Passarino--Veltman function $C_0$ is given by
\begin{equation}
C_0(s;m) = \frac{1}{2s} \ln^2 \left( \frac{\sqrt{1-\frac{4 m^2}{s}} - 1}{\sqrt{1-\frac{4 m^2}{s}} + 1}\right) . 
\end{equation}
We have used Package-X~\cite{Patel:2015tea} to calculate the one-loop integrals. 
Some comments regarding the dark matter annihilation into gamma gamma are in order. The cross section is not velocity suppressed, and therefore the rate can be large. 
The effective coupling $Z_B \gamma \gamma$ relies on a non-vanishing axial coupling of the fermions running in the loop to the $Z_B$, and therefore vanishes for the SM quarks. 
This means that only the new fermions needed for anomaly cancellation define the annihilation into photons.

\begin{table}[t]
\caption{Cross sections in $\unit[10^{-30}]{cm^3 s^{-1}}$. \label{tab:benchmark2}}
\begin{center}
\begin{tabular}{cccccc}
\hline
Scenario & $\langle \sigma v \rangle_{\gamma \gamma}$ &  $\langle \sigma v \rangle_{Z \gamma}$ &  $\langle \sigma v \rangle_{h \gamma}$ & $\langle \sigma v \rangle_{b \bar{b}}$  \\
\hline \hline
I & $11.5$ & $1.1$ & $8.0\times 10^{-7}$ & $0.023$ \\
II & $0.021$ & $15.2$ & $6.1 \times 10^{-7}$ & $0.024$ \\
III  & $0.047$ & $0.001$& $1.3 \times 10^{-6}$ & $0.048$ \\
\hline
\end{tabular}
\end{center}
\end{table}

\begin{figure}[b]
\centering
\includegraphics[width=0.8\linewidth]{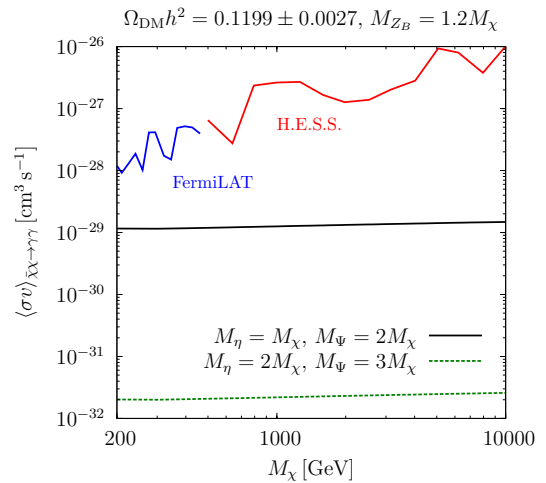}
\caption{Cross section for the DM annihilation into two gammas. We show two setups which have $M_{Z_B} = 1.2 M_\chi$. Setup 1 (black solid line) is defined by the parameter choice $M_\eta = M_\chi$, $M_\Psi = 2 M_\chi$. Setup 2 (green dashed line) is defined by the parameter choice $M_\eta = 2 M_\chi$, $M_\Psi = 3 M_\chi$. The gauge coupling $g_B$ is chosen such that in each point the dark matter relic density has the current value, $\Omega_\text{DM} h^2 = 0.1199 \pm 0.0027$~\cite{Ade:2013zuv}. Experimental limits on the cross section from Fermi-LAT~\cite{Ackermann:2015lka} and H.E.S.S.~\cite{Abramowski:2013ax} are given. \label{fig:indirectdetection}}
\end{figure}

In Table~\ref{tab:benchmark2}, we give the annihilation cross section to two gammas for the three benchmark scenarios defined in Table~\ref{tab:benchmark1}, together with the other annihilation cross sections for indirect detection. 
The prediction of the cross section for the annihilation into gamma gamma is given in Fig.~\ref{fig:indirectdetection}, together with the experimental bounds from Fermi-LAT~\cite{Ackermann:2015lka} and H.E.S.S.~\cite{Abramowski:2013ax}. As one can appreciate, one can achieve a rate close to the experimental limits when the new fermions are light.
The dark matter annihilation into a gamma and the SM Higgs, $\bar{\chi} \chi \to h \gamma$, is velocity suppressed. See Ref.~\cite{Duerr:2015wfa} for the details. 

%%%%%%%%%%%%%%%%%%%%%%%%%%%%%%%%%%%%%%%%%%%%%%%%%%%%%%%%%%%%%%%%%%%%%%%%%%%%%%%
\section{Gamma Spectrum}
%%%%%%%%%%%%%%%%%%%%%%%%%%%%%%%%%%%%%%%%%%%%%%%%%%%%%%%%%%%%%%%%%%%%%%%%%%%%%%%
The key feature of this type of model that allows for a visible gamma line is the fact that the DM annihilation into two photons is unsuppressed, whereas the tree-level annihilation to SM fermions and the final state radiation are suppressed. Generically the cross section times velocity for the final state radiation will be of the form
\begin{equation}\label{eq:FSR}
 \langle \sigma v \rangle_\text{FSR} = A \frac{M_q^2}{M_\chi^2} + B v^2 + {\cal O}(v^4),
\end{equation}
with some coefficients $A$ and $B$. Notice that the first term is suppressed by the quark masses, and the second term by the small DM velocity today. 

\begin{figure}[t]
 \centering
 \includegraphics[width=.8\linewidth]{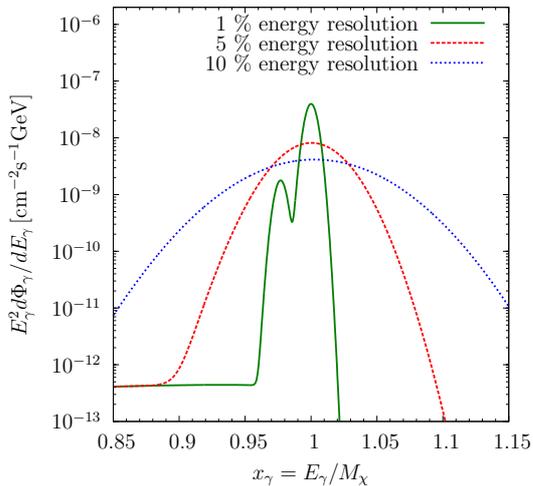}
 \caption{Differential spectrum of the annihilation of DM into gamma rays for benchmark scenario I. The three curves show the spectra with experimental resolutions of $10\%$ (blue dotted), $5\%$ (red dashed) and $1\%$ (green solid) in energy. We use the $J$-factor for the R3 region-of-interest by the Fermi-LAT collaboration, $J_\text{ann} = \unit[13.9\times 10^{22}]{GeV^2 cm^{-5}}$~\cite{Ackermann:2013uma}. The $\gamma \gamma$ and $Z \gamma$ lines are included. The continuum spectrum consists of the final state radiation, since the other channels leading to a continuum gamma spectrum are velocity suppressed. \label{fig:spectrum}}
\end{figure}

The differential gamma flux from the annihilation into two photons is given by
\begin{equation}
 \frac{d \Phi_\gamma}{d E_\gamma} = \frac{J_\text{ann} \langle \sigma v \rangle_{\gamma\gamma}}{4\pi M_\chi^2} G(E_\gamma, \xi/\omega, M_\chi),
\end{equation}
where we take the gamma line as a delta peak and model the detector resolution by a Gaussian
\begin{equation}
 G(E_\gamma, \xi/\omega,E_0) = \frac{1}{\sqrt{2 \pi} E_0 (\xi/\omega)}\text{e}^{-\frac{(E_\gamma - E_0)^2}{2 E_0^2 (\xi/\omega)^2}}.
\end{equation}
Here the parameter $\xi$ is a measure of the energy resolution of the detector and will vary typically between 1\% and 10\%, and the factor $\omega = 2 \sqrt{2 \log 2} \approx 2.35$ determines the full width at half maximum, with the standard deviation given by $\sigma_0 = E_0 \xi/\omega$. The $J$-factor $J_\text{ann}$ is the integral over the DM profile, which contains the astrophysical assumptions about the DM distribution in the galaxy and thus all astrophysical uncertainties. 

In Fig.~\ref{fig:spectrum} we show the differential spectrum for the dark matter annihilation into gamma rays in the benchmark scenario I in Table~\ref{tab:benchmark1}. The dominant contribution to the continuum in this scenario is FSR from the annihilation into $b$ quarks. Since the maximal energy of the FSR photons is given by
\begin{equation}
E_\gamma^\text{FSR,max} = M_\chi \left( 1- M_q^2/M_\chi^2 \right) , 
\end{equation}
the endpoint of the FSR gamma spectrum from the top quark will be far from the gamma line unless the DM is much heavier than the top quark. Then, however, the FSR cross section, see Eq.~\eqref{eq:FSR}, will be helicity suppressed by $M_t^2/M_\chi^2 \ll 1$.

Notice that the parameter choice in Fig.~\ref{fig:spectrum} is represented by the black line in Fig.~\ref{fig:indirectdetection} where the predicted cross section is close to the experimental bounds. Future experimental searches for gamma lines are expected to have very good energy resolution, see for example the proposed GAMMA-400~\cite{Topchiev:2015wva}, and one could hope to test this model in the foreseeable future. 

%%%%%%%%%%%%%%%%%%%%%%%%%%%%%%%%%%%%%%%%%%%%%%%%%%%%%%%%%%%%%%%%%%%%%%%%%%%%%%%
\section{Summary} 
%%%%%%%%%%%%%%%%%%%%%%%%%%%%%%%%%%%%%%%%%%%%%%%%%%%%%%%%%%%%%%%%%%%%%%%%%%%%%%%
In this letter we have shown that in simple models for Majorana dark matter one has a visible gamma line which could be used 
as a test of these dark matter models. In this context the dark matter is charged under a new gauge symmetry and the tree level annihilation into Standard Model particles is velocity suppressed. The contribution to the gamma continuum spectrum from dark matter annihilation is suppressed and the gamma line is distinguishable from the continuum. 

Our numerical results are in agreement with the bounds on the relic density, the direct detection limits and bounds from missing energy and leptophobic $Z^\prime$ searches at the LHC. In models with gauged baryon number and a self-conjugate dark matter candidate a striking line signature is a generic feature. Our novel results point toward an ideal scenario where using complementary experiments one can test a dark matter model.

%%%%%%%%%%%%%%%%%%%%%%%%%%%%%%%%%%%%%%%%%%%%%%%%%%%%%%%%%%%%%%%%%%%%%%%%%%%%%%%
\section*{Acknowledgments} 
%%%%%%%%%%%%%%%%%%%%%%%%%%%%%%%%%%%%%%%%%%%%%%%%%%%%%%%%%%%%%%%%%%%%%%%%%%%%%%%
We would like to thank H.\ H.\ Patel and  F.\ S.\ Queiroz for discussions. 

%%%%%%%%%%%%%%%%%%%%%%%%%%%%%%%%%%%%%%%%%%%%%%%%%%%%%%%%%%%%%%%%%%%%%%%%%%%%%%%

%%%%%%%%%%%%%%%%%%%%%%%%%%%%%%%%%%%%%%%%%%%%%%%%%%%%%%%%%%%%%%%%%%%%%%%%%%%%%%%


\begin{thebibliography}{99}
%%%%%%%%%%%%%%%%%%%%%%%%%%%%%%%%%%%%%%%%%%%%%%%%%%%%%%%%%%%%%%%%%%%%%%%%%%%%%%%

%\cite{Bringmann:2012ez}
\bibitem{Bringmann:2012ez}
  T.~Bringmann and C.~Weniger,
  ``Gamma Ray Signals from Dark Matter: Concepts, Status and Prospects,''
  Phys.\ Dark Univ.\  {\bf 1} (2012) 194
  [\href{http://www.arxiv.org/abs/arXiv:1208.5481}{arXiv:1208.5481 [hep-ph]}].
  %%CITATION = ARXIV:1208.5481;%%

%\cite{Gustafsson:2007pc}
\bibitem{Gustafsson:2007pc} 
  M.~Gustafsson, E.~Lundstrom, L.~Bergstrom and J.~Edsjo,
  ``Significant Gamma Lines from Inert Higgs Dark Matter,''
  Phys.\ Rev.\ Lett.\  {\bf 99} (2007) 041301
  [\href{http://www.arxiv.org/abs/astro-ph/0703512}{arXiv:astro-ph/0703512}].
  %%CITATION = ASTRO-PH/0703512;%%
  
%\cite{Duerr:2015mva}
\bibitem{Duerr:2015mva}
  M.~Duerr, P.~Fileviez Perez and J.~Smirnov,
  ``Scalar Singlet Dark Matter and Gamma Lines,''
  Phys.\ Lett.\ B {\bf 751} (2015) 119
  [\href{http://www.arxiv.org/abs/arXiv:1508.04418}{arXiv:1508.04418 [hep-ph]}].
  %%CITATION = doi:10.1016/j.physletb.2015.10.034;%%  
  
%\cite{Duerr:2015aka}
\bibitem{Duerr:2015aka}
  M.~Duerr, P.~Fileviez Perez and J.~Smirnov,
  ``Scalar Dark Matter: Direct vs. Indirect Detection,''
  \href{http://www.arxiv.org/abs/arXiv:1509.04282}{arXiv:1509.04282 [hep-ph]}.
  %%CITATION = ARXIV:1509.04282;%%
  
%\cite{Jungman:1995df}
\bibitem{Jungman:1995df}
  G.~Jungman, M.~Kamionkowski and K.~Griest,
  ``Supersymmetric dark matter,''
  Phys.\ Rept.\  {\bf 267} (1996) 195
  [\href{http://www.arxiv.org/abs/hep-ph/9506380}{arXiv:hep-ph/9506380}].
  %%CITATION = HEP-PH/9506380;%%
  
%\cite{Jackson:2009kg}
\bibitem{Jackson:2009kg} 
  C.~B.~Jackson, G.~Servant, G.~Shaughnessy, T.~M.~P.~Tait and M.~Taoso,
  ``Higgs in Space!,''
  JCAP {\bf 1004} (2010) 004
  [\href{http://www.arxiv.org/abs/arXiv:0912.0004}{arXiv:0912.0004 [hep-ph]}].
  %%CITATION = ARXIV:0912.0004;%%
  
%\cite{Arina:2009uq}
\bibitem{Arina:2009uq} 
  C.~Arina, T.~Hambye, A.~Ibarra and C.~Weniger,
  ``Intense Gamma-Ray Lines from Hidden Vector Dark Matter Decay,''
  JCAP {\bf 1003} (2010) 024
  [\href{http://www.arxiv.org/abs/arXiv:0912.4496}{arXiv:0912.4496 [hep-ph]}].
  %%CITATION = ARXIV:0912.4496;%%
  
%\cite{Bringmann:2012vr}
\bibitem{Bringmann:2012vr} 
  T.~Bringmann, X.~Huang, A.~Ibarra, S.~Vogl and C.~Weniger,
  ``Fermi LAT Search for Internal Bremsstrahlung Signatures from Dark Matter Annihilation,''
  JCAP {\bf 1207} (2012) 054
  [\href{http://www.arxiv.org/abs/arXiv:1203.1312}{arXiv:1203.1312 [hep-ph]}].
  %%CITATION = ARXIV:1203.1312;%%
  
%\cite{Dudas:2012pb}
\bibitem{Dudas:2012pb} 
  E.~Dudas, Y.~Mambrini, S.~Pokorski and A.~Romagnoni,
  ``Extra U(1) as natural source of a monochromatic gamma ray line,''
  JHEP {\bf 1210} (2012) 123
  [\href{http://www.arxiv.org/abs/arXiv:1205.1520}{arXiv:1205.1520 [hep-ph]}].
  %%CITATION = ARXIV:1205.1520;%%
  
%\cite{Giacchino:2013bta}
\bibitem{Giacchino:2013bta} 
  F.~Giacchino, L.~Lopez-Honorez and M.~H.~G.~Tytgat,
  ``Scalar Dark Matter Models with Significant Internal Bremsstrahlung,''
  JCAP {\bf 1310} (2013) 025
  [\href{http://www.arxiv.org/abs/arXiv:1307.6480}{arXiv:1307.6480 [hep-ph]}].
  %%CITATION = ARXIV:1307.6480;%%

%\cite{Kopp:2014tsa}
\bibitem{Kopp:2014tsa} 
  J.~Kopp, L.~Michaels and J.~Smirnov,
  ``Loopy Constraints on Leptophilic Dark Matter and Internal Bremsstrahlung,''
  JCAP {\bf 1404} (2014) 022
  [\href{http://www.arxiv.org/abs/arXiv:1401.6457}{arXiv:1401.6457 [hep-ph]}].
  %%CITATION = ARXIV:1401.6457;%%
  
%\cite{Coogan:2015xla}
\bibitem{Coogan:2015xla}
  A.~Coogan, S.~Profumo and W.~Shepherd,
  ``Monochromatic Gamma Rays from Dark Matter Annihilation to Leptons,''
  JHEP {\bf 1508} (2015) 074
  [\href{http://www.arxiv.org/abs/arXiv:1504.05187}{arXiv:1504.05187 [hep-ph]}].
  %%CITATION = doi:10.1007/JHEP08(2015)074;%%
  
%\cite{Duerr:2015wfa}
\bibitem{Duerr:2015wfa}
  M.~Duerr, P.~Fileviez Perez and J.~Smirnov,
  ``Simplified Dirac Dark Matter Models and Gamma-Ray Lines,''
  Phys.\ Rev.\ D {\bf 92} (2015) 083521
  [\href{http://www.arxiv.org/abs/arXiv:1506.05107}{arXiv:1506.05107 [hep-ph]}].
  %%CITATION = doi:10.1103/PhysRevD.92.083521;%%  
  
%\cite{Garcia-Cely:2015dda}
\bibitem{Garcia-Cely:2015dda}
  C.~Garcia-Cely, A.~Ibarra, A.~S.~Lamperstorfer and M.~H.~G.~Tytgat,
  ``Gamma-rays from Heavy Minimal Dark Matter,''
  JCAP {\bf 1510} (2015) 058
  \href{http://www.arxiv.org/abs/arXiv:1507.05536}{arXiv:1507.05536 [hep-ph]}].
  %%CITATION = doi:10.1088/1475-7516/2015/10/058;%

 %\cite{Ishiwata:2011hr}
\bibitem{Ishiwata:2011hr}
  K.~Ishiwata and M.~B.~Wise,
  ``Higgs Properties and Fourth Generation Leptons,''
  Phys.\ Rev.\ D {\bf 84} (2011) 055025
 [\href{http://www.arxiv.org/abs/arXiv:1107.1490}{arXiv:1107.1490 [hep-ph]}].
  %%CITATION = ARXIV:1107.1490;%%
  
%\cite{Duerr:2013dza}
\bibitem{Duerr:2013dza}
  M.~Duerr, P.~Fileviez Perez and M.~B.~Wise,
  ``Gauge Theory for Baryon and Lepton Numbers with Leptoquarks,''
  Phys.\ Rev.\ Lett.\  {\bf 110} (2013) 231801.
  [\href{http://www.arxiv.org/abs/arXiv:1304.0576}{arXiv:1304.0576 [hep-ph]}].
  %%CITATION = ARXIV:1304.0576;%%

%\cite{Perez:2014qfa}
\bibitem{Perez:2014qfa}
  P.~Fileviez Perez, S.~Ohmer and H.~H.~Patel,
  ``Minimal Theory for Lepto-Baryons,''
  Phys.\ Lett.\ B {\bf 735} (2014) 283
  [\href{http://www.arxiv.org/abs/arXiv:1403.8029}{arXiv:1403.8029 [hep-ph]}].
  %%CITATION = ARXIV:1403.8029;%%

%\cite{Duerr:2013lka}
\bibitem{Duerr:2013lka}
  M.~Duerr and P.~Fileviez Perez,
  ``Baryonic Dark Matter,''
  Phys.\ Lett.\ B {\bf 732} (2014) 101
  [\href{http://www.arxiv.org/abs/arXiv:1309.3970}{arXiv:1309.3970 [hep-ph]}].
  %%CITATION = ARXIV:1309.3970;%%
  
%\cite{Duerr:2014wra}
\bibitem{Duerr:2014wra}
  M.~Duerr and P.~Fileviez Perez,
  ``Theory for Baryon Number and Dark Matter at the LHC,''
  Phys.\ Rev.\ D {\bf 91} (2015)  095001
  [\href{http://www.arxiv.org/abs/arXiv:1409.8165}{arXiv:1409.8165 [hep-ph]}].
  %%CITATION = ARXIV:1409.8165;%%
   
%\cite{Ohmer:2015lxa}
\bibitem{Ohmer:2015lxa}
  S.~Ohmer and H.~H.~Patel,
  ``Leptobaryons as Majorana Dark Matter,''
  Phys.\ Rev.\ D {\bf 92} (2015) 055020
  [\href{http://www.arxiv.org/abs/arXiv:1506.00954}{arXiv:1506.00954 [hep-ph]}].
  %%CITATION = ARXIV:1506.00954;%%

  %\cite{Gondolo:1990dk}
\bibitem{Gondolo:1990dk}
  P.~Gondolo and G.~Gelmini,
  ``Cosmic abundances of stable particles: Improved analysis,''
  Nucl.\ Phys.\ B {\bf 360} (1991) 145.
  %%CITATION = NUPHA,B360,145;%%
  
%\cite{Ade:2013zuv}
\bibitem{Ade:2013zuv}
  P.~A.~R.~Ade {\it et al.} [Planck Collaboration],
  ``Planck 2013 results. XVI. Cosmological parameters,''
  Astron.\ Astrophys.\  {\bf 571} (2014) A16
  [\href{http://www.arxiv.org/abs/arXiv:1303.5076}{arXiv:1303.5076 [astro-ph.CO]}].
  %%CITATION = ARXIV:1303.5076;%%
  
%\cite{Buchmueller:2014yoa}
\bibitem{Buchmueller:2014yoa}
  O.~Buchmueller, M.~J.~Dolan, S.~A.~Malik and C.~McCabe,
  ``Characterising dark matter searches at colliders and direct detection experiments: Vector mediators,''
  JHEP {\bf 1501} (2015) 037
  [\href{http://www.arxiv.org/abs/arXiv:1407.8257}{arXiv:1407.8257 [hep-ph]}].
  %%CITATION = ARXIV:1407.8257;%%

%\cite{Patel:2015tea}
\bibitem{Patel:2015tea}
  H.~H.~Patel,
  ``Package-X: A Mathematica package for the analytic calculation of one-loop integrals,''
  Comput.\ Phys.\ Commun.\  {\bf 197} (2015) 276
  [\href{http://www.arxiv.org/abs/arXiv:1503.01469}{arXiv:1503.01469 [hep-ph]}].
  %%CITATION = ARXIV:1503.01469;%%
  

%\cite{Ackermann:2015lka}
\bibitem{Ackermann:2015lka}
  M.~Ackermann {\it et al.} [Fermi-LAT Collaboration],
  ``Updated search for spectral lines from Galactic dark matter interactions with pass 8 data from the Fermi Large Area Telescope,''
  Phys.\ Rev.\ D {\bf 91} (2015) 122002
  [\href{http://www.arxiv.org/abs/arXiv:1506.00013}{arXiv:1506.00013 [astro-ph.HE]}].
  %%CITATION = ARXIV:1506.00013;%%
 
%\cite{Abramowski:2013ax}
\bibitem{Abramowski:2013ax}
  A.~Abramowski {\it et al.} [HESS Collaboration],
  ``Search for Photon-Linelike Signatures from Dark Matter Annihilations with H.E.S.S.,''
  Phys.\ Rev.\ Lett.\  {\bf 110} (2013) 041301
  [\href{http://www.arxiv.org/abs/arXiv:1301.1173}{arXiv:1301.1173 [astro-ph.HE]}].
  %%CITATION = ARXIV:1301.1173;%%
  
%\cite{Ackermann:2013uma}
\bibitem{Ackermann:2013uma}
  M.~Ackermann {\it et al.} [Fermi-LAT Collaboration],
  ``Search for gamma-ray spectral lines with the Fermi large area telescope and dark matter implications,''
  Phys.\ Rev.\ D {\bf 88} (2013) 082002
  [\href{http://www.arxiv.org/abs/arXiv:1305.5597}{arXiv:1305.5597 [astro-ph.HE]}].
  %%CITATION = ARXIV:1305.5597;%%
  
%\cite{Topchiev:2015wva}
\bibitem{Topchiev:2015wva}
  N.~P.~Topchiev {\it et al.},
  ``GAMMA-400 gamma-ray observatory,''
  \href{http://www.arxiv.org/abs/arXiv:1507.06246}{arXiv:1507.06246 [astro-ph.IM]}.
  %%CITATION = ARXIV:1507.06246;%%
  
%%%%%%%%%%%%%%%%%%%%%%%%%%%%%%%%%%%%%%%%%%%%%%%%%%%%%%%%%%%%%%%%%%%%%%%%%%%%%%%
\end{thebibliography}
\end{document}